\documentclass[aps,pre,onecolumn,notitlepage,twoside, amsmath,amssymb,aps,bbm,floatfix,superscriptaddress]{revtex4-1}

\usepackage{graphicx}
\usepackage{dcolumn}
\usepackage{bm}
\usepackage{xcolor}

\newcommand{\be}{\begin{equation}}
\newcommand{\ee}{\end{equation}}
\newcommand{\bea}{\begin{eqnarray}}
\newcommand{\eea}{\end{eqnarray}}

\usepackage{mathtools}

\begin{document}

\title{Solving the subset sum problem with a nonideal biological computer}

\author{Michael Konopik}
 \affiliation{Institute for Theoretical Physics I, University of Stuttgart, D-70550 Stuttgart, Germany}
\author{Till Korten}
 \affiliation{B CUBE - Center for Molecular Bioengineering, Technische Universit\"at Dresden, 01069 Dresden, Germany}
\author{Heiner Linke}
\affiliation{NanoLund and Solid State Physics, Lund University, S-22100 Lund, Sweden}
\author{Eric Lutz}
 \affiliation{Institute for Theoretical Physics I, University of Stuttgart, D-70550 Stuttgart, Germany}

\begin{abstract} 
We consider the solution of the subset sum problem based on a parallel  computer consisting of self-propelled biological agents moving in a nanostructured network that encodes the NP-complete task in its geometry. We develop an approximate analytical method to analyze the effects of small errors in the nonideal junctions composing the computing network by using a Gaussian confidence
interval approximation of the multinomial distribution. We concretely evaluate the probability distribution for error-induced paths and determine the minimal number of agents required to obtain a proper solution. We finally validate our theoretical results with exact numerical simulations of the subset sum problem for different set sizes and error probabilities.

\end{abstract}

\maketitle

\noindent \textbf{\large 1. Introduction}\\

\noindent
Solving complex problems requires high-performance computing methods. Since the number of computing steps typically increases exponentially with the size of a problem, multi-processor parallel algorithms are better suited to solve such tasks than  single-processor serial algorithms \cite{gra03,pac07}. In addition to traditional electronic computers,  massively parallel computers have recently been realized using biological systems, such as DNA computing {\cite{ade94,lip95,ouy97,mao00,bra02,qia11,erl17}} and network-based biocomputing \cite{nic06,nic16}, which use small DNA molecules or cytoskeletal filaments as processors. These biological agents  can be mass-produced at low cost  and  added to the computation in amounts matching the problem size. As a result, such biological architectures are easily highly parallelizable.

\begin{figure}[t]
\includegraphics[width=0.75\textwidth]{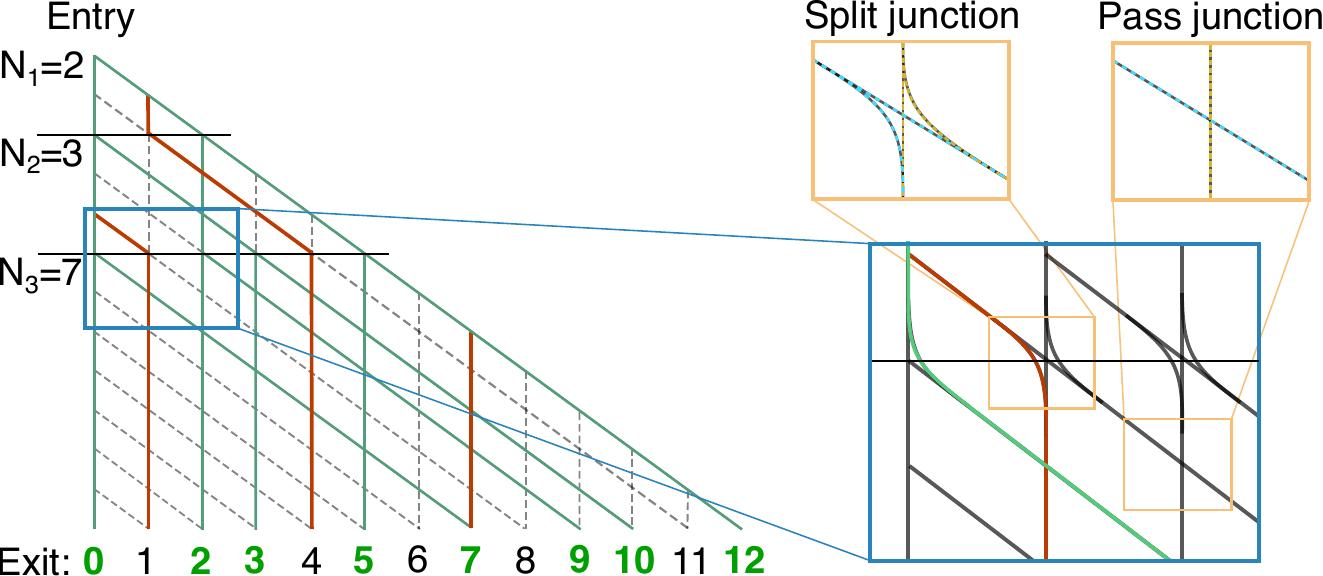}
\caption{Network used to solve the subset sum problem. Self-propelled motor proteins enter the grid at the top. They move vertically down at a pass junction and diagonally down at a split junction.  The exit numbers at the bottom correspond to potential solutions subset sum problem (target sums $T$). Correct  results for the set \{2, 3, 7\} are indicated in green boldface numbers. {Green lines are examples of correct paths, while red lines show examples of incorrect ones.}}
\label{f1}
 \end{figure}

We here consider a nondeterministic-polynomial-time complete (NP-complete) problem, known as the subset sum problem \cite{gar79,kle05}: given a set of positive integers and an integer $T$ (the target sum), the general question is to determine whether  there is a subset whose sum is exactly $T$.  A proof-of-principle solution of this problem for the set $\{2,5,9\}$ has been provided  using   molecular-motor-propelled agents moving  in a nanostructured network that encodes the combinatorial problem into its geometry  \cite{nic16}. The network consists of a grid of channels  which can be traversed by cytoskeletal filaments from top (entry point) to bottom (exit points) (figure~\ref{f1}). The grid is made of   two types of junctions: pass junctions that allow an agent to continue on its previous path and split junctions that allow an agent to switch lanes, or not, with equal probability. Travelling vertically down the grid  corresponds to adding 0 to the final total number, while moving  diagonally corresponds to adding 1. By properly positioning the split junctions,  the target sums can be determined from the position of the agents at the bottom of the grid. {The network is explored stochastically by individual agents. Therefore, depending on the desired confidence level, a sufficient total number of agents must be supplied in order to obtain a significant result.}

In the following, we develop a general  approach based on a Gaussian confidence
interval approximation of the multinomial distribution \cite{wal13} that allows   to  analytically analyze the effects of  small errors  in nonideal  junctions. We evaluate, in particular,  the probability distribution for error-induced paths, taking into account  errors both in pass and split junctions, and introduce {a procedure with which the minimal number of agents required to obtain a correct solution of the problem may be determined}.  Both issues are essential for the successful experimental implementation of the {biological computation} strategy \cite{Heiner}. We finally compare our theoretical results with exact numerical simulations of the subset sum problem for different set sizes and error probabilities.\\

\noindent \textbf{\large 2. Necessary number of agents for ideal junctions}\\

\noindent The protein filaments semi-randomly traverse the grid and are independent of  each other. While it is clear that one has to use at least $N_\text{sol}=2^s$ agents for a set of $s$ integers, $\{N_1,N_2,...\}$, using this number is quite unlikely to result in a complete solution of the problem. In fact, the larger the problem size, the exponentially more unlikely it is that this number of agents would result in a complete solution. We first estimate  the smallest number of agents needed to solve the combinatorial problem with a sufficiently strong confidence, by assuming ideal split junctions (with a split ratio of 50\%) and ideal pass junctions (the agents are guaranteed to keep on moving along their vertical/diagonal path). 
Under these conditions, all the paths resulting in a valid number of the $N_\text{sol}$ solutions are equally likely. Without loss of generality, we  assume that all solutions are unique. There are hence  $N_\text{sol}$ paths that one agent may choose from.

 This multinomial problem may be solved by finding the formally exact statistics using combinatorial methods. However, the  exact statistics may  be formulated only indirectly (confidence intervals for the exits have to be chosen such that at least one agent exits through each of them) and solved numerically. But as all possibilities have to be computed, this becomes  rapidly intractable for large $s$. In addition,  exploring all possible solutions is equivalent to solving the problem itself and, thus,  makes the biological computer obsolete. We here employ another approach that approximates the problem using normal distributions \cite{wal13}. The advantage of this method  is that it can be applied regardless of the size $s$ of the set, and be improved, if higher accuracy is needed, by going beyond the Gaussian approximation \cite{wal13}.

We denote by $x_i$ the number of agents in outcome $i$, $p_i$ the corresponding probability (with $\sum_i p_i=1$) and  $N= \sum_{i}^{N_\text{sol}}x_i$ the total number of agents ($p_i = 1/N_\text{sol}$ for ideal junctions). The multinomial distribution for $k$ different outcomes  and $N$ independent trials is then given by \cite{seb11}, \begin{equation}
P(k,N)=\frac{N!}{x_1! x_2!\cdots x_k!} p_1^{x_1} p_2^{x_2}\cdots p_k^{x_k}.
\label{Multin}
\end{equation}
This distribution has  mean $\bar{x}_i= N p_i$ and standard deviation $\sigma_i = \sqrt{N p_i q_i}$ with $q_i=1-p_i$ \cite{seb11}.

\begin{figure}[t]
\includegraphics[width=0.7\textwidth]{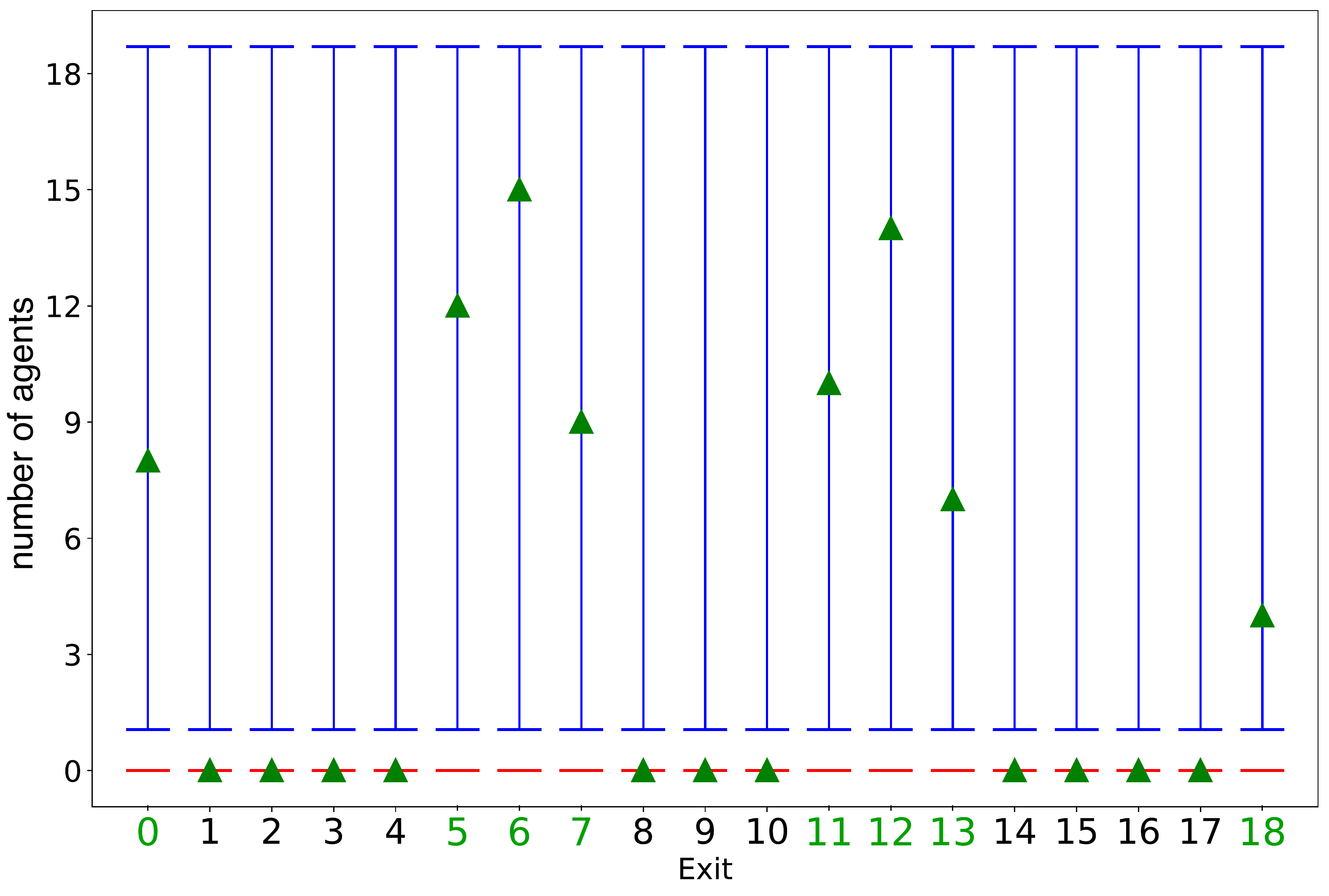}
\caption{Number of agents leaving an ideal grid for the set $\{5,6,7\}$. Green triangles represent the  number of numerically simulated agents ending at the respective exit. {The theoretical description of the stochastic computation is indicated by the} $3\sigma$ confidence intervals based on a Gaussian approximation of the  multinomial distribution {(see discussion around \eqref{perfecttotalagents})}, shown in blue for each possible exit. According to \eqref{numberOfAgentsPerfect}, the minimal number of agents is $N_\text{min}=79$ with   $n_i=1$ in this case. {This condition guarantees that each correct exit has a finite number of agents: each  green triangle is then  either  in a blue (correct) confidence interval or in a red (incorrect) confidence interval, allowing to effectively distinguish the two outcomes (red confidence intervals have  vanishing sizes for the ideal grid considered here)}.}
\label{f2}
 \end{figure}

Since  agents traverse the grid randomly, one can only state a confidence interval (which quantifies the margin of error) in which a given number of agents will use any of the possible paths. There are a variety of confidence interval approximations for the multinomial distribution \cite{wal13}: the simplest is the normal approximation interval, which approximates the distribution by a Gaussian.  The symmetric confidence interval  is given in this case by $[\bar{ x}_i - \ell \sigma_i, \bar{ x}_i + \ell \sigma_i]$ \cite{wal13}. The $\ell=3$ confidence interval for the normal distribution guarantees that the number of agents traversing through the respective paths lies in this interval with a probability of $p_\text{single-success}\approx 99\%$ for individual outcomes. The success probability of a full computation is accordingly given by  $p_{\text{success}}= p_\text{single-success}^{N_\text{sol}}$. 
If we now demand at least $n_i\geq 1$ agents per path (in order to  find a proper solution to the subset sum problem) then the lower boundary of the confidence interval gives a determining equation for the necessary total number of agents $N_\text{min}$,
\begin{equation}
n_i = \bar{x}_i - 3 \sigma_i= N_\text{min} p_i - 3 \sqrt{N_\text{min} p_i q_i}.
\label{perfecttotalagents}
\end{equation}
Solving this equation for $N_\text{min}$ results in
\begin{equation}
N_\text{min}= \frac{3}{2 p_i}\left(3 q_i + \frac{2}{3} n_i + \sqrt{9 q_i^2 + 4 q_i n_i}\right).
\label{numberOfAgentsPerfect}
\end{equation}
Using  \eqref{numberOfAgentsPerfect} with $n_i$=1 guarantees with $99\%$ certainty that there is at least one agent {per correct path. The corresponding success probability of the full computation will then be $p_{\text{success}} =0.92, 0.85, 0.72, 0.53$ for the set sizes $s=3,4,5,6$. The success probability  may be further improved by taking a value $\ell>3$. For concreteness, we will consider the case $\ell=3$  in the remainder.}  Figure~\ref{f2} shows  the exits of the various numbers from 0 to the sum of all numbers for the set $\{2,3,7\}$ -- green triangles indicate the number of simulated agents ending at the respective exits. In this case, the minimum of agents  is $N_\text{min}=79$. A number is a solution of the subset sum problem if the arriving number of agents lies within the (blue) confidence interval. Since the setup is assumed to be ideal, {the exits will have  agents going through them only  valid paths. Therefore, the error confidence intervals (red)  are centered around 0 and have length 0 in this case. Any correct exit will have agents in a blue confidence interval.}   The corresponding minimal numbers for agents for $s=4$ and $s= 5$ are respectively $N_\text{min} = 166$ and  $N_\text{min}= 340 $ (for $n_i=1$).\\

\noindent \textbf{\large 3. Effects of errors for nonideal junctions}\\

\noindent The assumptions of an ideal  grid do not {necessarily} apply to an actual experiment. The pass junctions {may} indeed not {be} perfectly able to keep  agents on their path and  split junctions {may} not {be} perfectly symmetric. These errors may result in incorrect or missing solutions  of the subset sum problem. We next analyze the effects of small errors both in the pass junctions and in the split junctions on the the minimal number of agents $N_\text{min}$.

Errors in  pass junctions mean that  agents take new paths that are not actual solutions, since they act then as split junctions. On the other hand, faulty split junctions cause an imbalance of valid paths, as the split probability deviates from 50\%. 
Errors generated by nonideal split junctions  may  be simply accommodated for by taking the smallest probability of the unlikeliest path for the number estimation of equation \eqref{numberOfAgentsPerfect}. This gives a good estimate as long as the probability is not too low ($p_i N\gtrsim5$) to guarantee  the validity of the Gaussian approximation \cite{bar89}. 

By contrast, errors induced by pass junctions  are  more complex to treat and result in the creation of new paths. The total number of split junctions is given by the cardinality of the considered set, $N_\text{SJ}=s$. The total number of pass junctions is equal to the sum of all numbers in the set, $N_\text{tot}=\sum_i N_i $, minus the number of split junctions, $N_\text{PJ} = N_\text{tot}- N_\text{SJ}$. Assuming that a pass junction has an error probability of $p_\text{PJ}$,  the probability of an agent passing through the grid without any pass error may then be evaluated as $p_\text{PJ}^{\text{c}}=(1- p_\text{PJ})^{N_\text{PJ}}=(1- p_\text{PJ})^{N_\text{tot}-N_\text{SJ}}$.  Errors therefore depend  exponentially  on the size of the problem. Even the simplest grid of just one number can be highly incorrect if the chosen number is too large in comparison to the error probability. Qualitatively speaking, pass junction errors have to be as small as possible to allow the biological computer to work for larger problems. 

Quantitatively, the nonideal system can be modeled as being comprised of two parts: the correct part that has to be done by at least $N_\text{min}$  agents as given by equation \eqref{numberOfAgentsPerfect} and the incorrect part {$N_{\text{FP}}$} that results in a number of (noisy) agents wrongly added in the various exits. This model may  be described with the help of  a binomial distribution: agents either take a correct path with a given probability, or a wrong one. Thus, the minimal number of agents $N_\text{min}^\text{non}$ that have to be used in a nonideal grid, such that there are  $N_\text{min}$ agents that traverse it correctly, can be calculated  similarly as done before with the multinomial distribution. We obtain (using the same 3$\sigma$ confidence interval),
\begin{equation}
\label{4}
N_\text{min}^\text{non} = \frac{3}{2 p_\text{PJ}^{\text{c}}} \left(3 q_\text{PJ}^{\text{c}}+ \frac{2}{3}N_\text{min} + \sqrt{9 p_\text{PJ}^{\text{c 2}} + 4 q_\text{PJ}^{\text{c}} N_\text{min}}\right).
\end{equation}
The above number guarantees that there are enough agents to fulfill the computation condition of the ideal case \eqref{numberOfAgentsPerfect}. However, it does not yet incorporate the effect of agents taking faulty paths, whose number is $N_\text{FP}= N_\text{min}^\text{non}-N_\text{min}$. These agents  create  noise on the various exits that needs to be accounted for. In particular, agents originating from correct and incorrect paths ought to be distinguished in an effective manner.\\

\noindent \textbf{\large 3.1. Probability distribution for error-induced paths}\\

\noindent While outcomes were considered  {to be} equally probable in the case of an ideal grid, this is no longer true for nonideal, noisy  grids. Figure~\ref{f1} shows that agents can move through a large number of   possible paths before reaching a given exit point.
In particular, the central outcomes may be reached by more trajectories than the outer ones.  This suggests that the effect of errors  will generally be nonhomogeneous. We shall  in this Section evaluate the distribution of error-induced paths in the limit of small error probabilities. For larger error probabilities, the probability of successful propagation through the grid becomes negligibly small:  for a subset sum problem that contains 18 pass junctions, an error probability per pass junction of $p_\text{PJ} = 0.2$ will, for example, lead to a successful propagation probability of only $p_\text{PJ}^\text{c} = 0.018$.  {In these cases, where the correct agents arriving at each exit are solely a small perturbation of the wrong agents, either a more involved error estimation has to be done, or incorrect agents have to be registered and discarded from the counting to still being able to correctly do the stochastic computation.}

Pass and split junctions are  not independent of each other in general. For  wrong paths, a split junction indeed acts as a particularly bad pass junction, creating a stronger inhomogeneity in the noise outcomes.
Faulty split junctions also complicate the treatment significantly. However, their effect is expected to be weak  for small errors.   We concretely make the following  assumptions:
(1) the error probability is equal for all pass junctions,  (2)  the error probability  is related to a change of the  initial direction of an agent, and (3) split junctions, with the exception of the first one, do not contribute {strongly} to the shape of the (weak) noise {and may be incorporated in an approximative manner}. We will assess the range of validity of these hypotheses by comparison with an exact numerical evaluation in Sect. 4 below.
We will first  consider a simplistic grid consisting only of faulty pass junctions (and no split junctions) before accounting for the average effects of split junctions at the end of this Section.

Based on  the above assumptions, it is possible to find a recursive formula for the number of paths {$A_{i,N_\text{tot}}^m$} for each potential solution $i$ in dependence of the number of turns (defined as error-induced changes of paths) $m$. Using the compact notation $Z=N_\text{tot}$, we have,
\begin{equation}
A^1_{i,Z} = 
\begin{cases}
1 & \text{for } i = Z,0\\
0 & \text{else}
\end{cases}
, \quad A^2_{i,Z} = 
\begin{cases}
0 & \text{for } i = Z,0\\
2 & \text{else}
\end{cases}
\label{combe1}
\end{equation}
\begin{equation}
 A^3_{i,Z} = 
\begin{cases}
0 & \text{for } i = Z,0\\
(Z - 1 - i) + (i - 1) \equiv  {}_v A^3_{i,Z} + {}_d A^3_{i,Z}  & \text{else}
\end{cases}
\label{combe2}
\end{equation}
\begin{equation}
\begin{aligned}
 A^m_{i,Z} &= 
\begin{cases}
0 & \text{for } i = 0,...,g(m) \text{ or } n-g(m),...,n \\
\sum_{j=1}^{i-g(m)} {}_v A^{m-1}_{i-j,Z-j} + \sum_{j=1}^{n-(g(m-1) +1 + i)}  {}_d A^{m-1}_{i,Z-j}  & \text{else}
\end{cases}\\
&\equiv
\begin{cases}
0 & \text{for } i = 0,...,g(m) \text{ or } n-g(m),...,n \\
 {}_v A^m_{i,Z} + {}_d A^m_{i,Z}  & \text{else}\\
\end{cases}
\end{aligned}
\label{combe3}
\end{equation}
with the modulo function, $g(m) = (m-m \,\text{mod} 2)/2-1$,
which is needed due to the fact that there is an alternating pattern in the recursive calculation{, as will be explained in the following}. We have here defined the quantities ${}_v A^m_{i,Z}$   and  ${}_d A^m_{i,Z}$ that respectively describe contributions from vertical and diagonal paths. 

\begin{figure}
\centering
\includegraphics[width=0.49\textwidth]{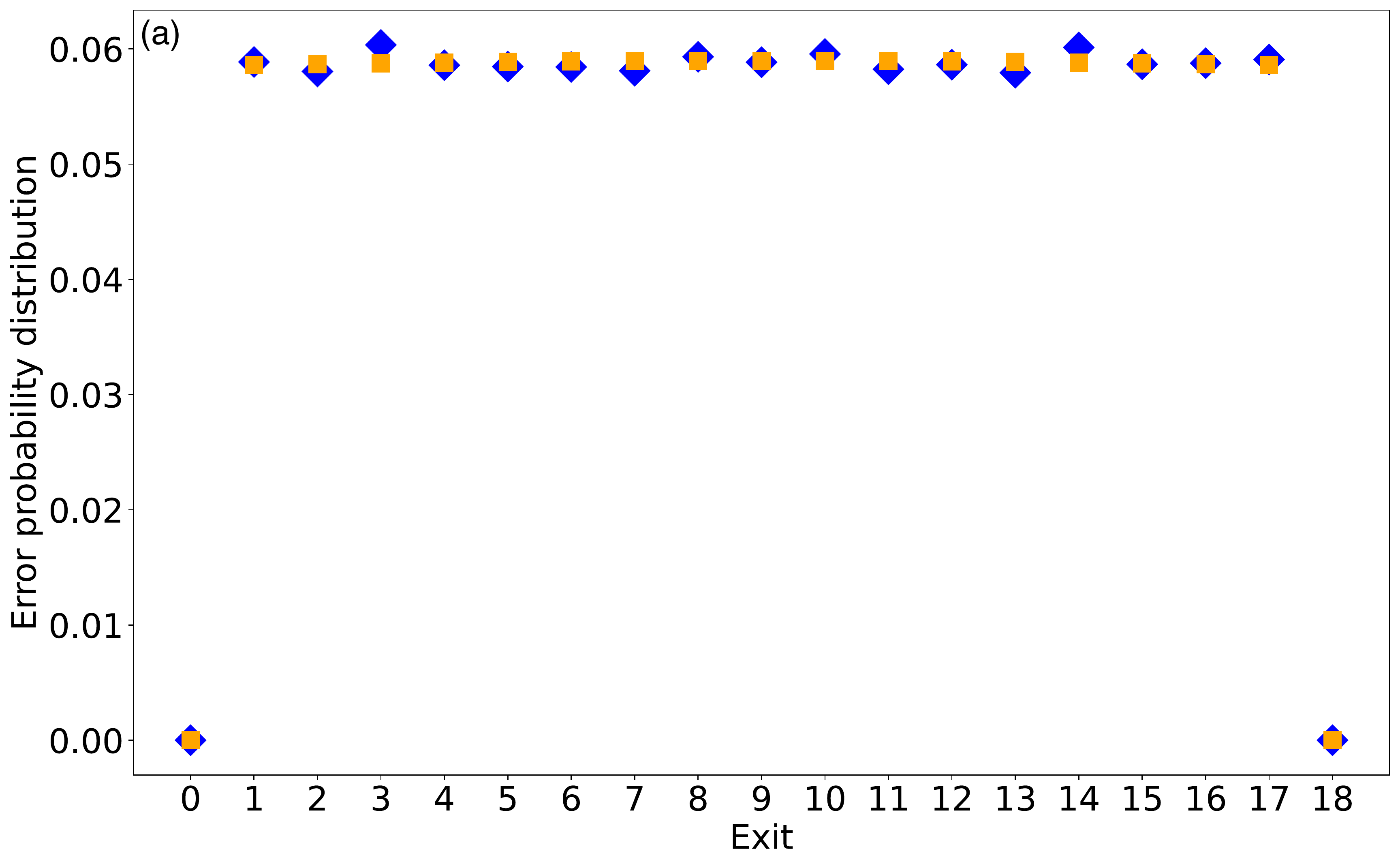}
\includegraphics[width=0.49\textwidth]{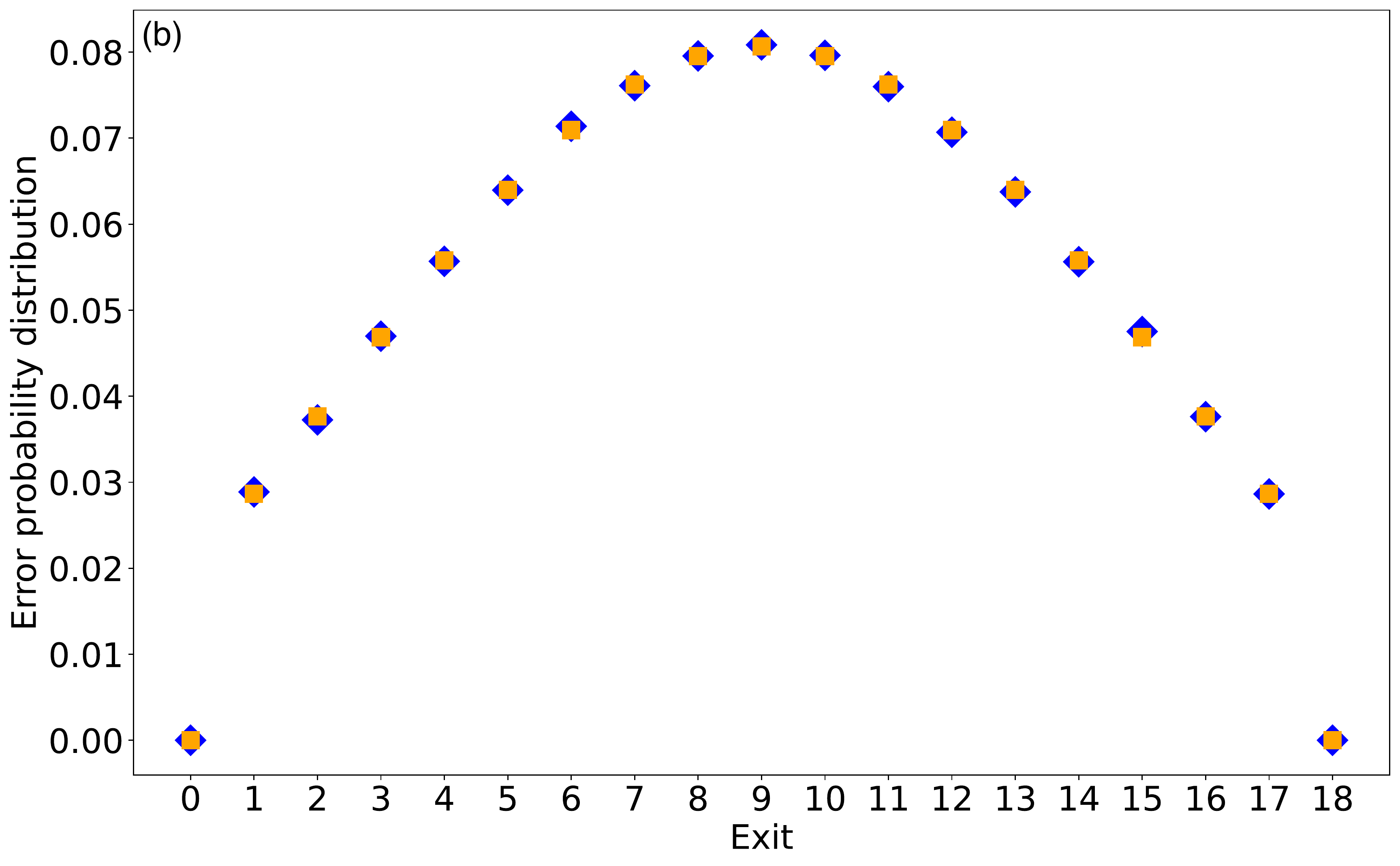}
\caption{Analytical probability distribution of error-induced outcomes $p_i^\text{non}$, \eqref{nonprob}, (orange squares) and corresponding numerical simulations (blue diamonds) for a  simplistic grid of length 18 consisting only of nonideal pass junctions, for (a) $p_\text{PJ}=0.01$ and (b) $p_\text{PJ}=0.15$.
The distribution is flat for small error probabilities and  peaked at the position of the central exit point for larger error probabilities, indicating that outcomes at the edges of the grid are less likely. There is excellent agreement between theory and simulations in both cases.}
\label{errornumericsvgl}
\end{figure}

To understand this recursive formula, it is  instructive to start with the case $m = 1$. There is no case $m=0$, since initially an agent has to choose between going down (and potentially ending up at exit 0 by keeping this direction), or moving right (and potentially arriving at exit $Z$ by again keeping this direction), which corresponds to $m=1$.  For $m = 2$, an agent has to do one detour while being on  either the (vertical) 0-path or the (diagonal) $Z$-path. It  therefore  cannot reach these two values but any other exit may be reached via a detour from both of them. As a result, there are  two possibilities for any potential solution not being either 0 or $Z$. This distinction between the vertical path $(v)$ and diagonal $(d)$ path is also used for the  general case of $m$ wrong turns. For $m=3$, a first turn on the 0- or $Z$-path reduces the grid size and one receives multiple $m=2$ contributions for the varying sizes, differing from each other by either starting vertically $v$ or diagonally $d$. For $m=3$ this still results in a flat distribution
 For $m > 3$, it is necessary to sum up all the possible lengths of the initial vertical or diagonal paths, for multiple recursive contributions, which yields  non-flat distributions.

The probability distribution of error-induced outcomes  follows accordingly as,\begin{equation}
p_{i}^\text{non} = \frac{1}{C_Z} \sum_{m=1}^{Z-1} A^m_{i,Z} (p_\text{PJ})^{m-1} (q_\text{PJ})^{Z-m}  p_\text{SJ}
\label{nonprob}
\end{equation}
with the normalization constant,
\begin{equation}
\label{nonprob1}
C_Z = \sum_{i=1}^{Z-1}\left[ \sum_{m=1}^{Z-1} A^m_{i,Z} (p_\text{PJ})^{m-1} (q_\text{PJ})^{Z-m} p_\text{SJ} \right].
\end{equation}
It is important to note that the probability distribution \eqref{nonprob} is only defined  for $i \neq (0, Z)$, since these  are actual solutions of the subset sum problem and cannot be reached via wrong trajectories. We will take an equal initial split junction, $p_\text{SJ}=1/2$, in the following.

 \begin{figure}
\centering
\includegraphics[width=0.49\textwidth]{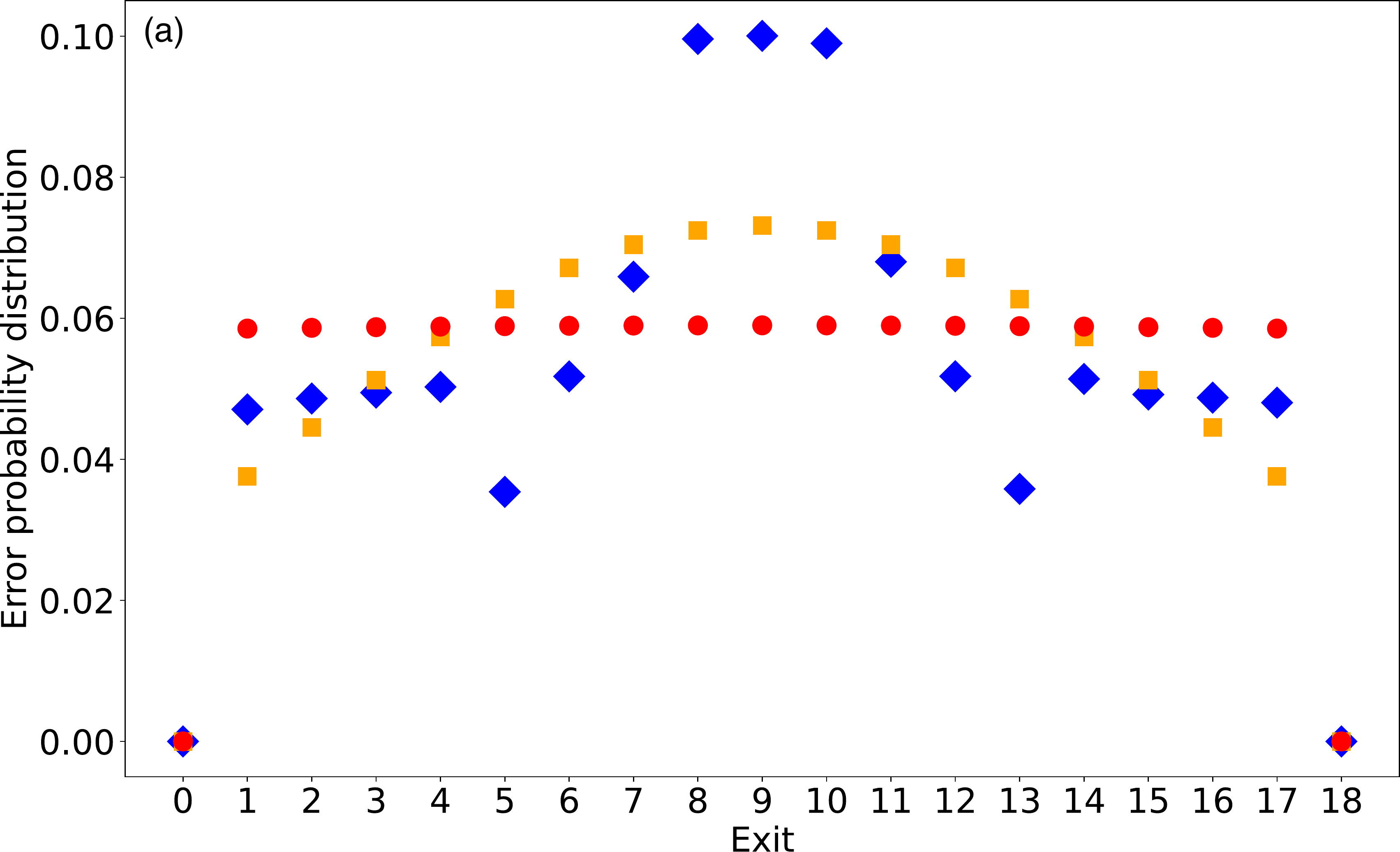}
\includegraphics[width=0.49\textwidth]{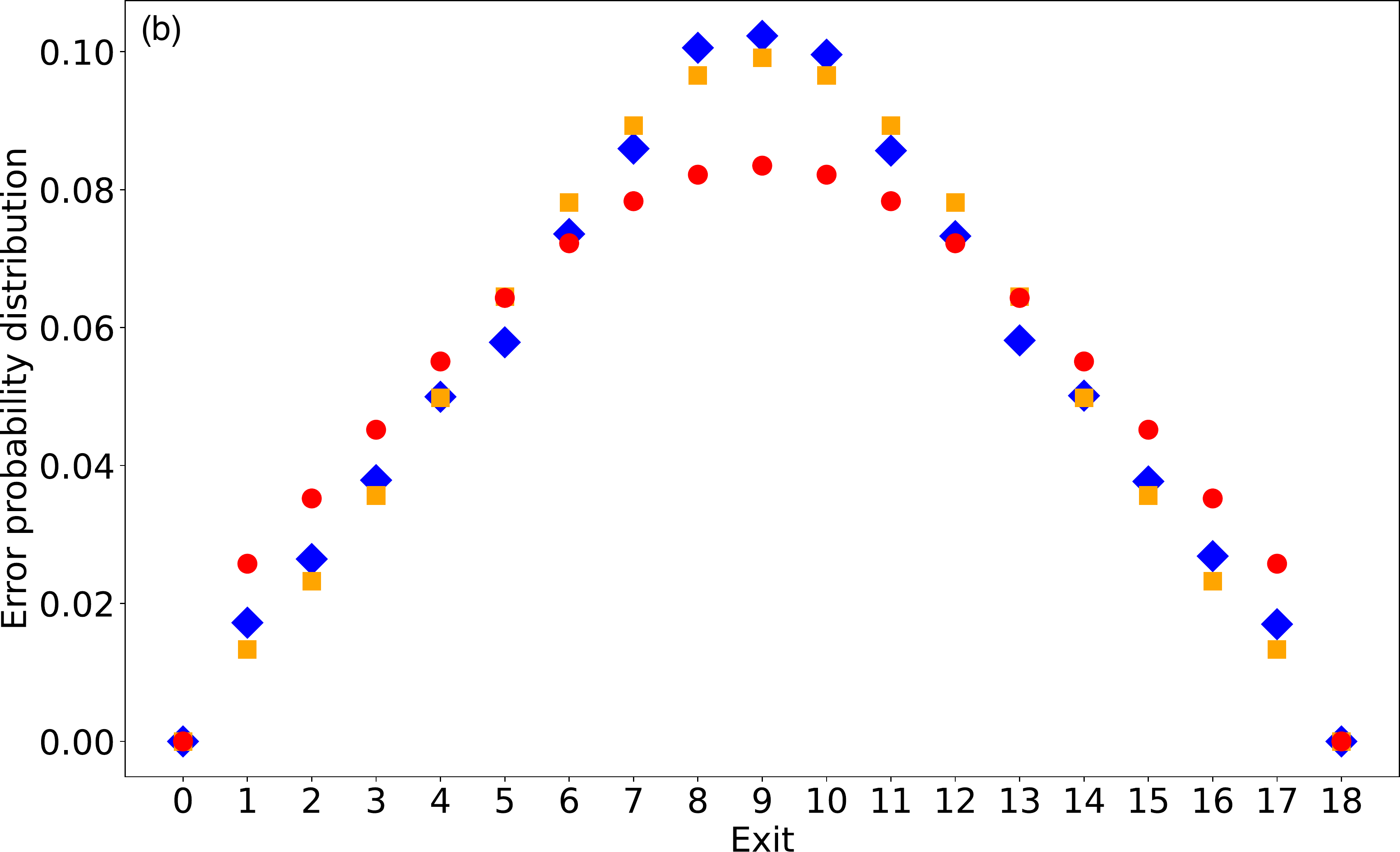}
\includegraphics[width=0.49\textwidth]{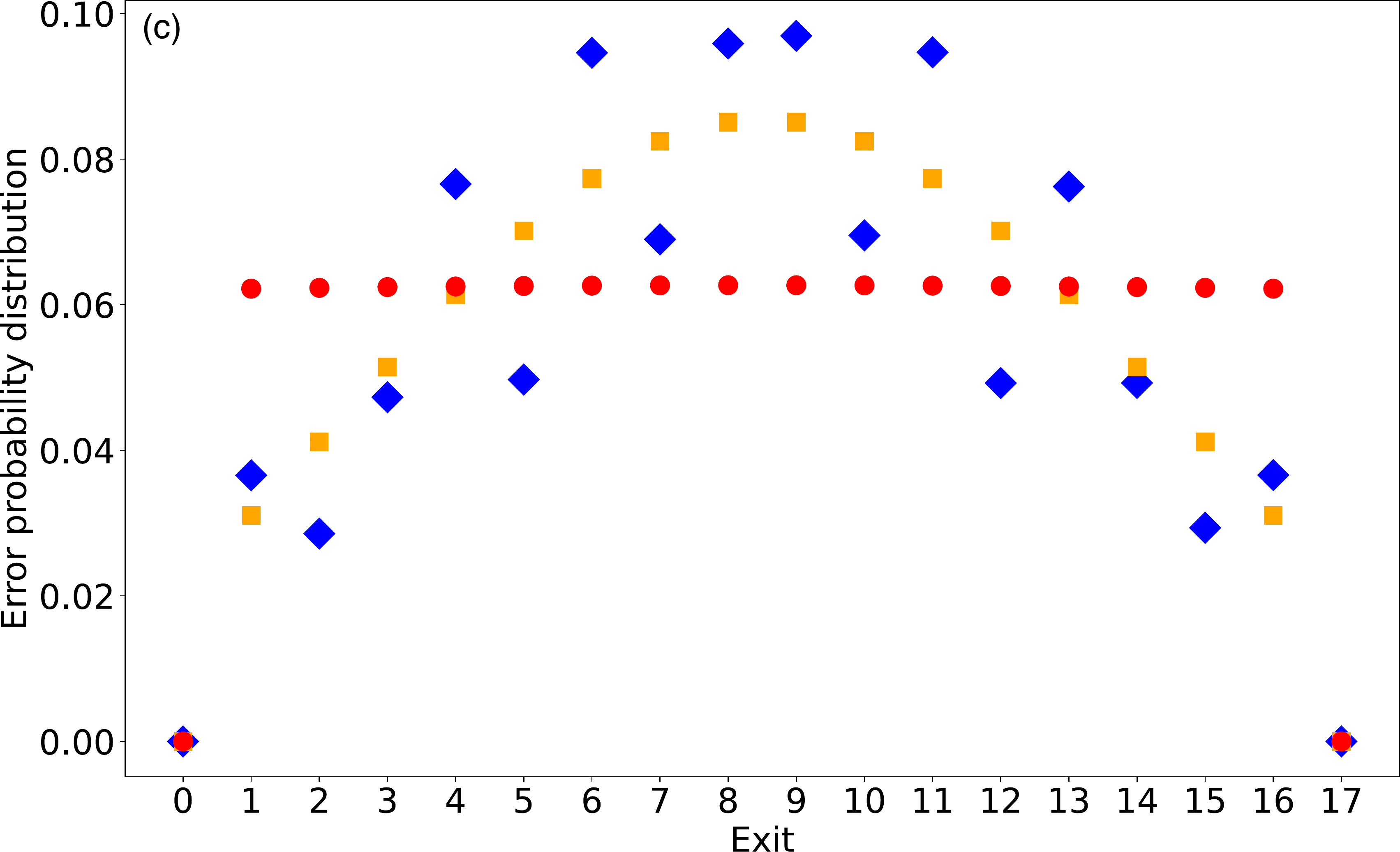}
\includegraphics[width=0.49\textwidth]{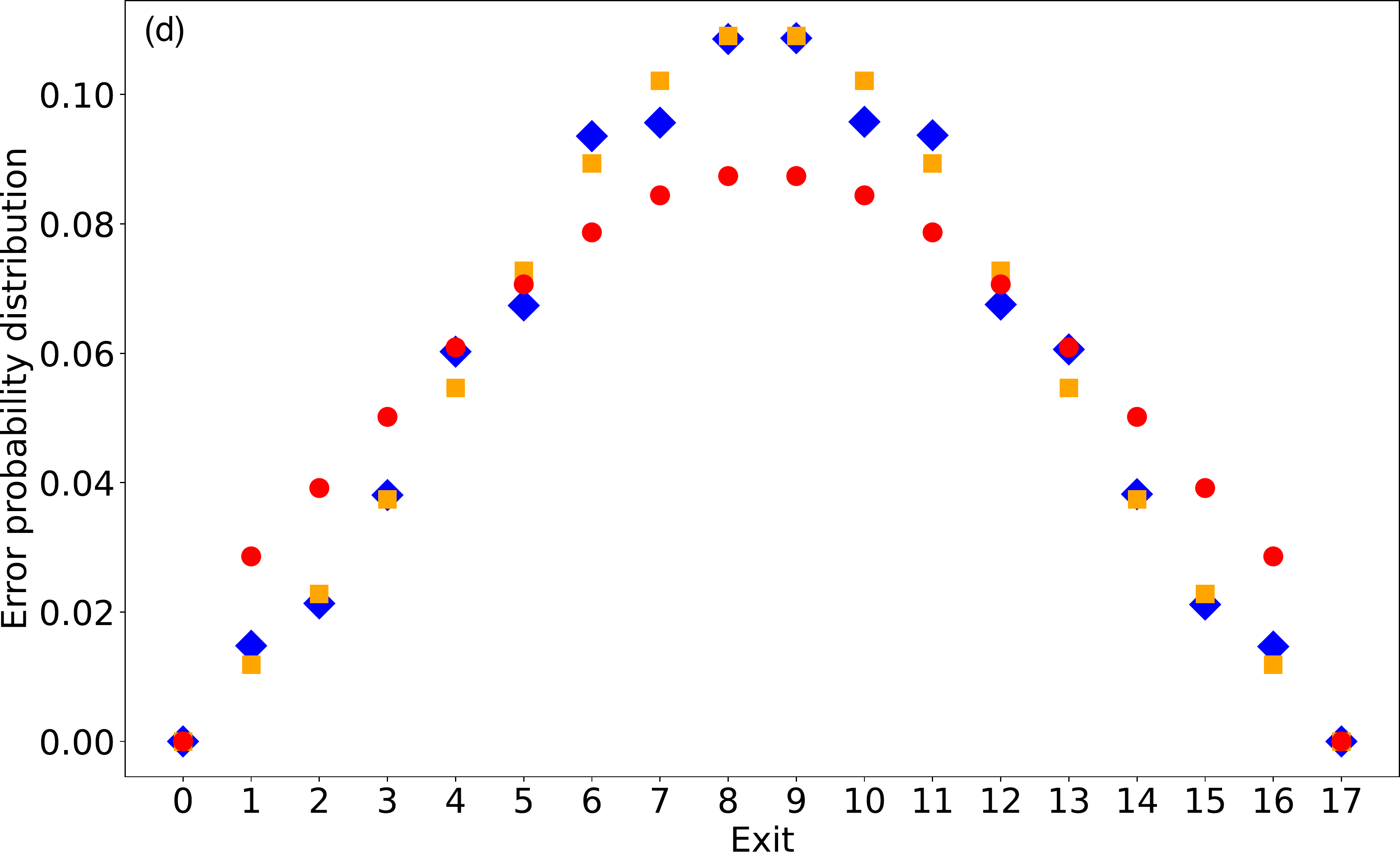}
\caption{Analytical probability distribution of error-induced outcomes $p_i^\text{non}$, \eqref{nonprob},  with nonideal pass junctions, {without taking  the interaction between split and pass junctions into account} (red circles), {taking the interaction between the two into  account via the effective error probability \eqref{nocheinp}} (orange squares), and corresponding exact numerical simulations (blue diamonds), for (a),(c) $p_\text{PJ}=0.01$ and (b),(d) $p_\text{PJ}=0.15$, for the set \{5,6,7\} (top) and $\{2,3,5,7\}$ (bottom). The approximate treatment of the split junctions via the effective probability \eqref{nocheinp} exhibits very good agreement with the exact simulations for larger error probabilities and larger set sizes, when the effect of the split junctions is better averaged.  
}
\label{errornumericsvglsplit}
\end{figure}

{We have so far considered pass junction errors for a setup only comprised of such junctions. However,  split junctions also exist  for  agents traversing the grid incorrectly. Considering  split junctions as particularly bad pass junctions,  $p_\text{PJ} \rightarrow  p_\text{SJ}$, one has to incorporate their effects for the effective error distribution.} 
 A simple approach to account for these effects is to use an effective error probability for the pure pass junction errors in~\eqref{nonprob} and \eqref{nonprob1}, by incorporating the split junction probability into all the pass junctions evenly: 
\begin{equation}
p_\text{PJ} = p_\text{eff} = 1- (1-p_\text{PJ}) \frac{(1-p_\text{SJ})^{s/Z}}{(1-p_\text{PJ}) ^{s/Z}} = 1-q_\text{PJ}.
\label{nocheinp}
\end{equation}
Equation \eqref{nocheinp} indicates that the existence of the split junctions enhances the effective error probability, even for small pass junction error probabilities (see Sect. 4 below). {It is important to emphasize that \eqref{nocheinp} only modifies the shape of the distribution of the errors, not how many agents propagate through the grid erroneously.}\\

\begin{figure}
  \centering
  \includegraphics[width=0.49\textwidth]{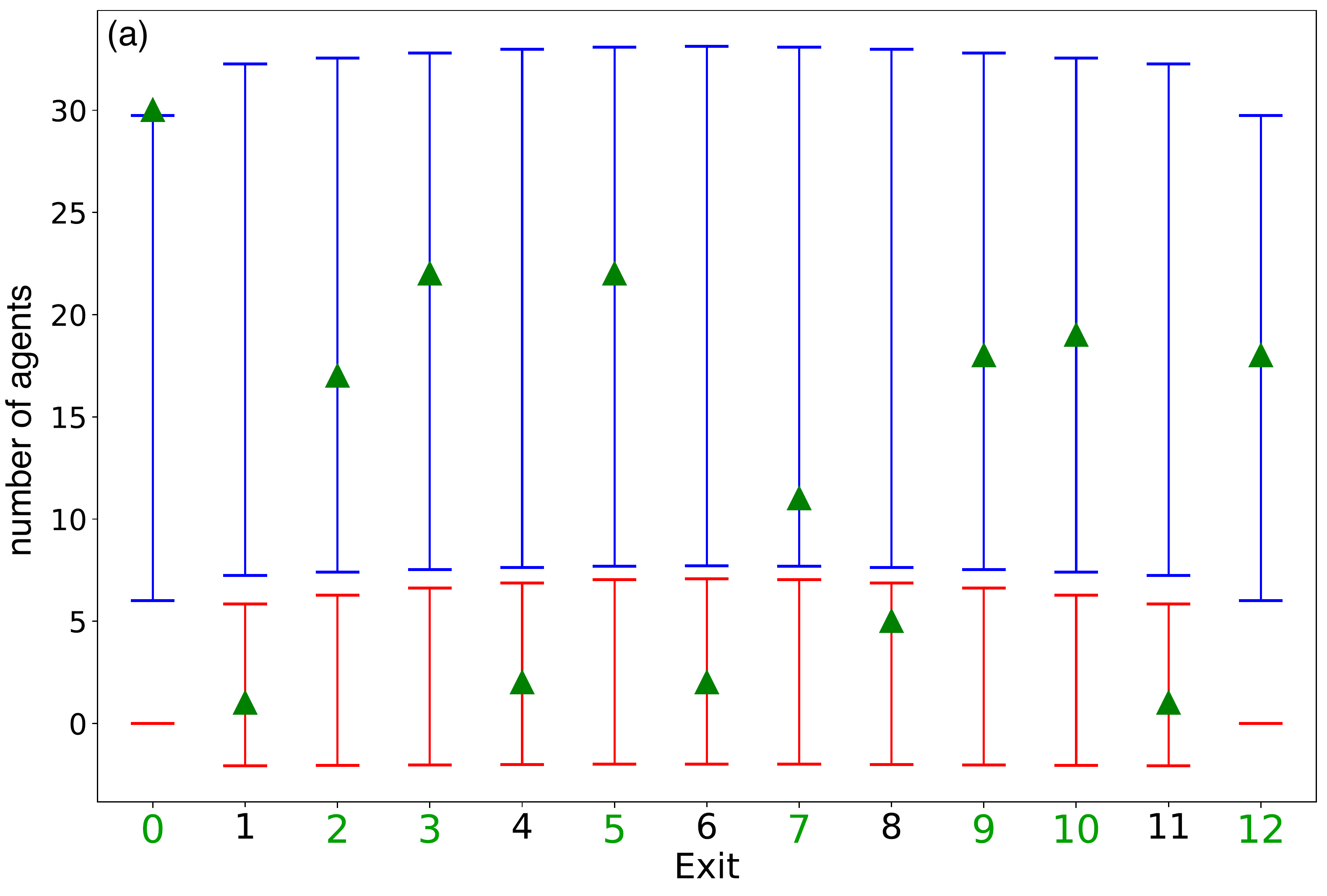}
\includegraphics[width=0.49\textwidth]{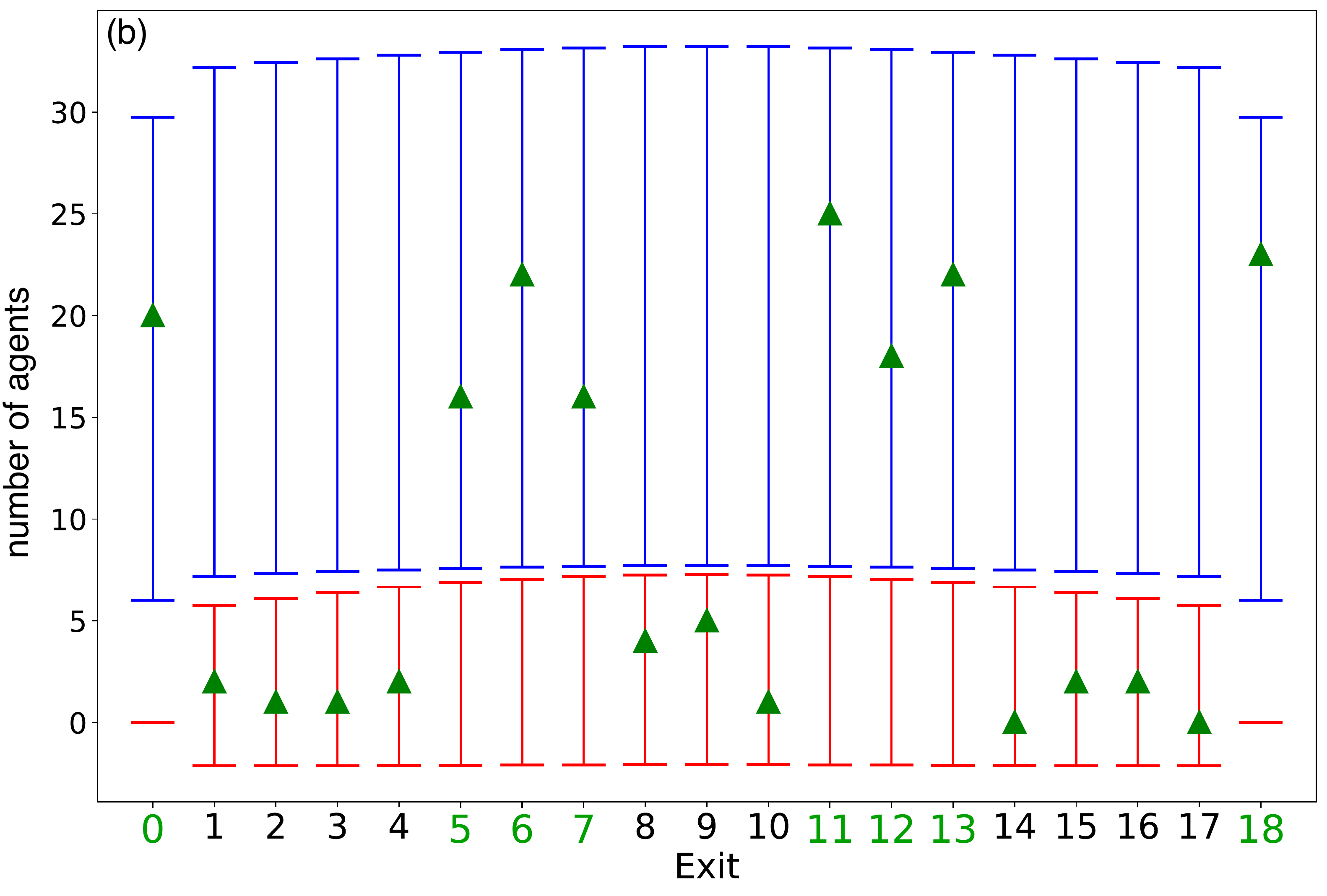}
  \includegraphics[width=0.49\textwidth]{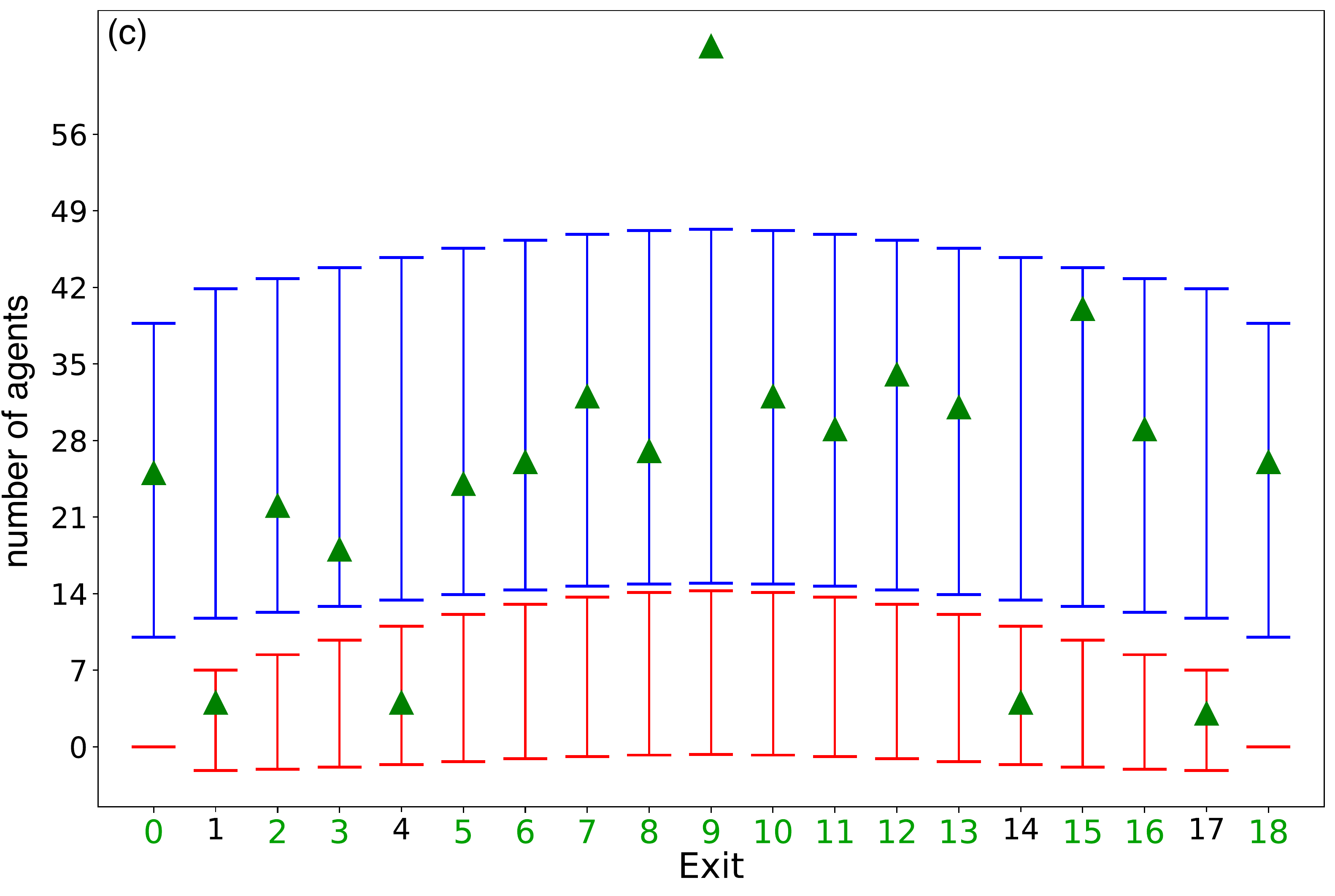}
\includegraphics[width=0.49\textwidth]{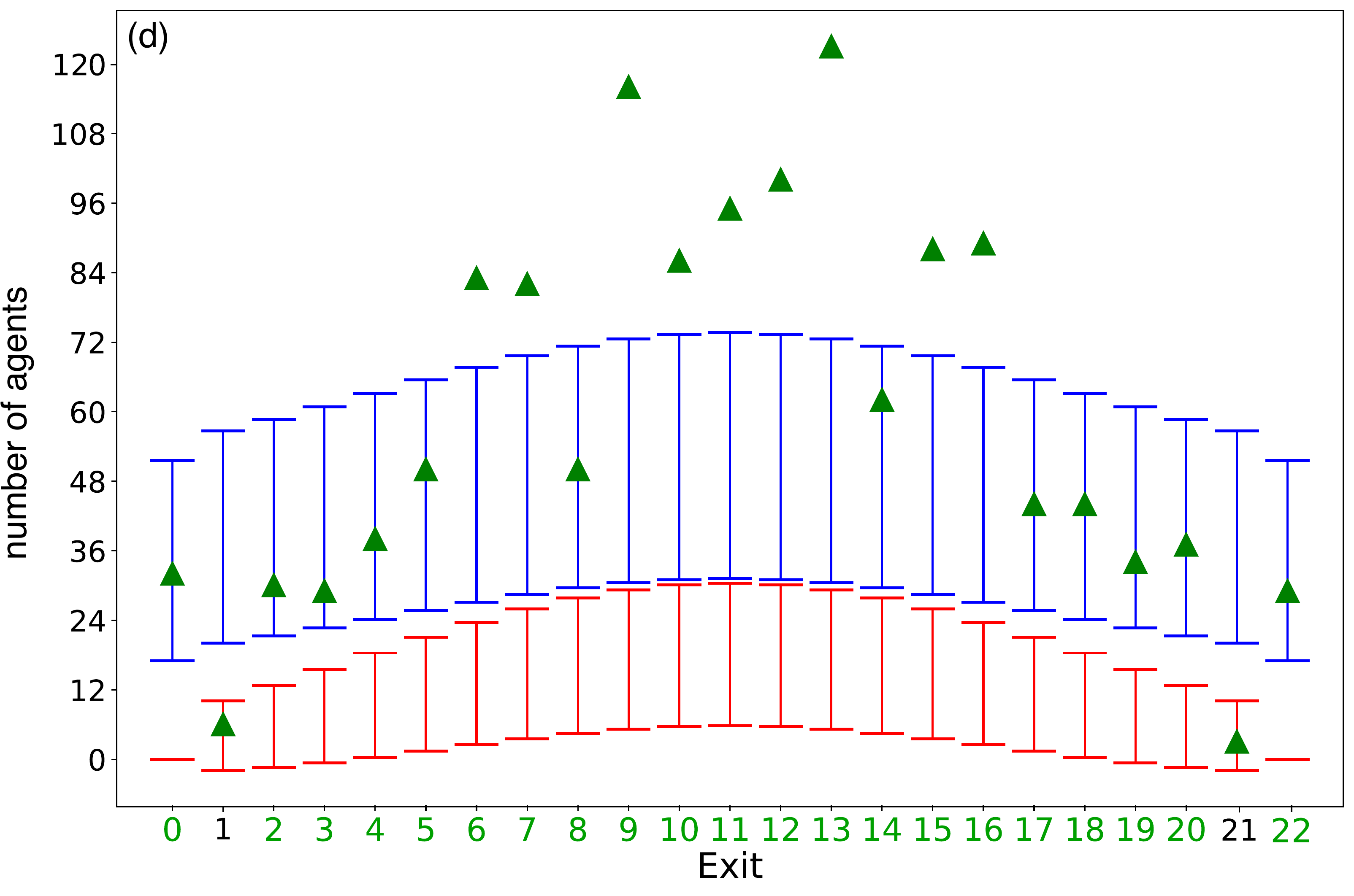}
\caption{Number of agents leaving a nonideal grid with an error probability $p_\text{PJ} = 0.01$, for various sets: $(a)$ $\{2,3,7\}$,  (b) $ \{5,6,7\}$, (c) $ \{2,3,6,7\}$  and  (d) $ \{2,3,7,6,4\}$. Compared with the ideal grid shown in figure~2, we here have nonzero (red) $3\sigma$ confidence intervals for wrong paths. The correct (blue)  intervals contain a stochastically independent sum of the confidence intervals of the theoretical description of the  correct and incorrect agents (as given by \eqref{perfecttotalagents} and \eqref{figciter}).  A large number of pass junctions increases the {effective error probability \eqref{nocheinp}, leading}  to a stronger curvature of the overall shape of the confidence intervals.  In all four cases, agents originating from correct and incorrect paths can be clearly distinguished: {their total number is either in a blue interval (correct exit) or in a red interval (incorrect exit)}. The exits 9 and 10 can be reached by two combinations of paths in (c) and thus are above the blue confidence intervals intervals. The same happens for exits located closed to the center point in (d).   The minimal number of agents increases from (a) $N_\text{min}^\text{non} = 168$,  (b) $N_\text{min}^\text{non} = 182$, (c) $N_\text{min}^\text{non}= 474$ to (d) $N_\text{min}^\text{non} = 1350$.}
\label{1percentvgl}
\end{figure}

\noindent \textbf{\large 3.2. Mean and standard deviation from the error probability distribution}\\

\noindent 
Having determined the probability distribution (\ref{nonprob}) of an individual faulty outcome $i$, we may now  evaluate mean and standard deviation for $N_\text{FP}$ faulty trials using the multinomial distribution as done in Sect. 2 for the ideal grid. Since we are interested in the minimal necessary number  of agents, the most relevant probability  is the one that has the largest possible deviations. This is  the case for the central outcome which corresponds to $i_\text{max}=Z/2$ or, more precisely, to $i_\text{max} = g(Z) + 1$, since this instance involves the largest number of possible paths. We obtain
\begin{equation}
\bar n_\text{max} = p_{i_\text{max}}^\text{non} N_\text{FP};~~~~\sigma_\text{max} = \sqrt{N_\text{FP} p_{i_\text{max}}^\text{non} q_{i_\text{max}}^\text{non} }.
\label{figciter}
\end{equation}
 According to this formula, it is possible to distinguish right from wrong paths  if the number of 
 agents $n_i$ per outcome $i$ obeys (taking again a 3$\sigma$ confidence interval),
 \begin{equation}
 n_i \geq p_{i_\text{max}}^\text{non} N_\text{FP} + 3 \sqrt{N_\text{FP} p_{i_\text{max}}^\text{non} q_{i_\text{max}}^\text{non} }.
 \label{rightwrongdistinction}
 \end{equation}
 A solution to the subset sum problem can be found if the total number of agents $N_\text{min}$ is such that equation \eqref{rightwrongdistinction} is satisfied. In this case, the measured signal will be either within the confidence interval of the error estimate or it will be above that value due to the additional agents coming from the correct paths. As a consequence, confidence intervals for purely wrong outcomes, and those for the sum of correct and wrong outcomes (corresponding to the stochastically independent sum of their respective average and variance) will not overlap. Expression \eqref{rightwrongdistinction} generalizes the condition $n_i\geq 1$ of the ideal grid. For nonideal grids, $n_i$ will in general grow with the error probability $p_\text{PJ}$. \\
 
%
%
 \begin{figure}
  \centering
  \includegraphics[width=0.49\textwidth]{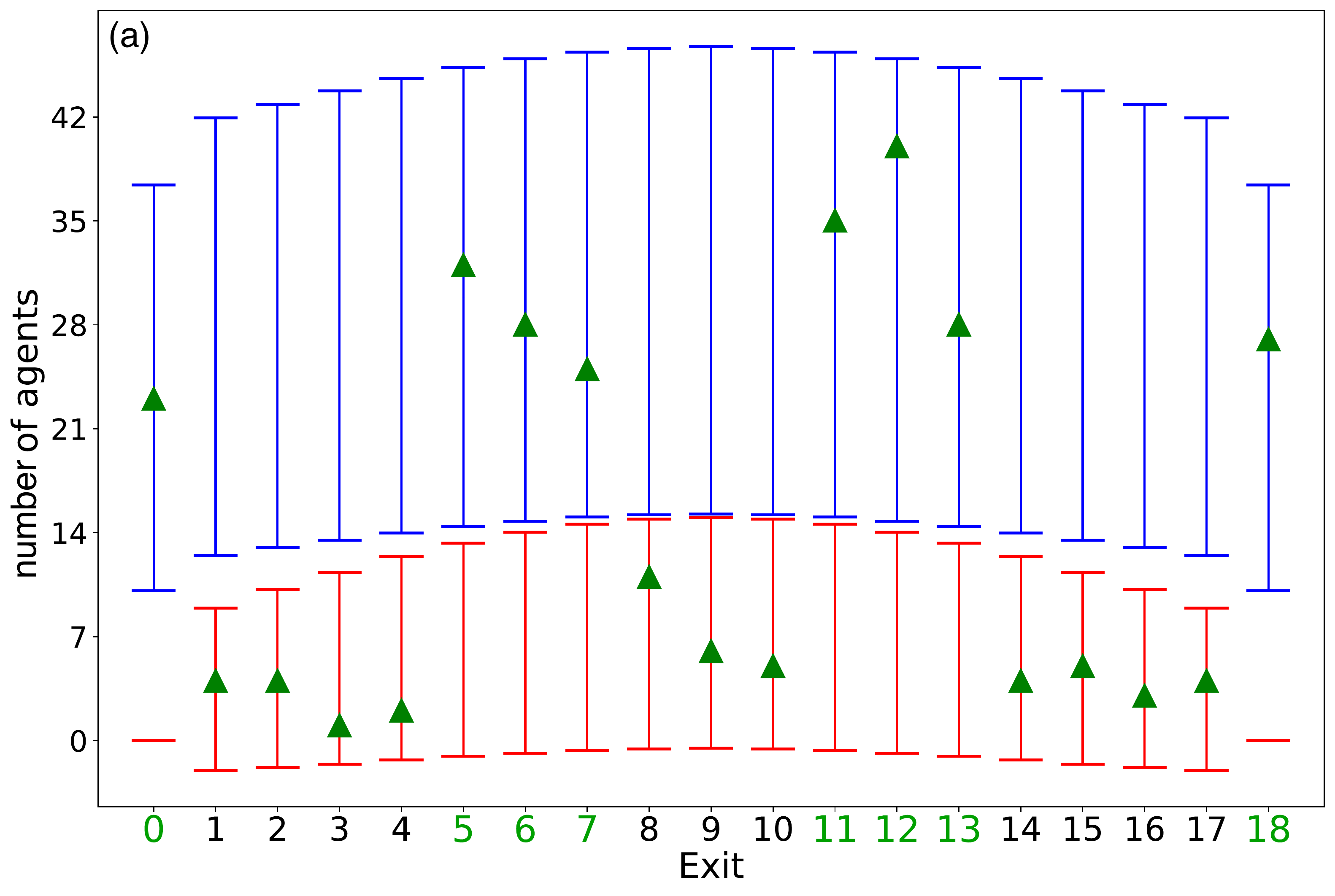}
\includegraphics[width=0.49\textwidth]{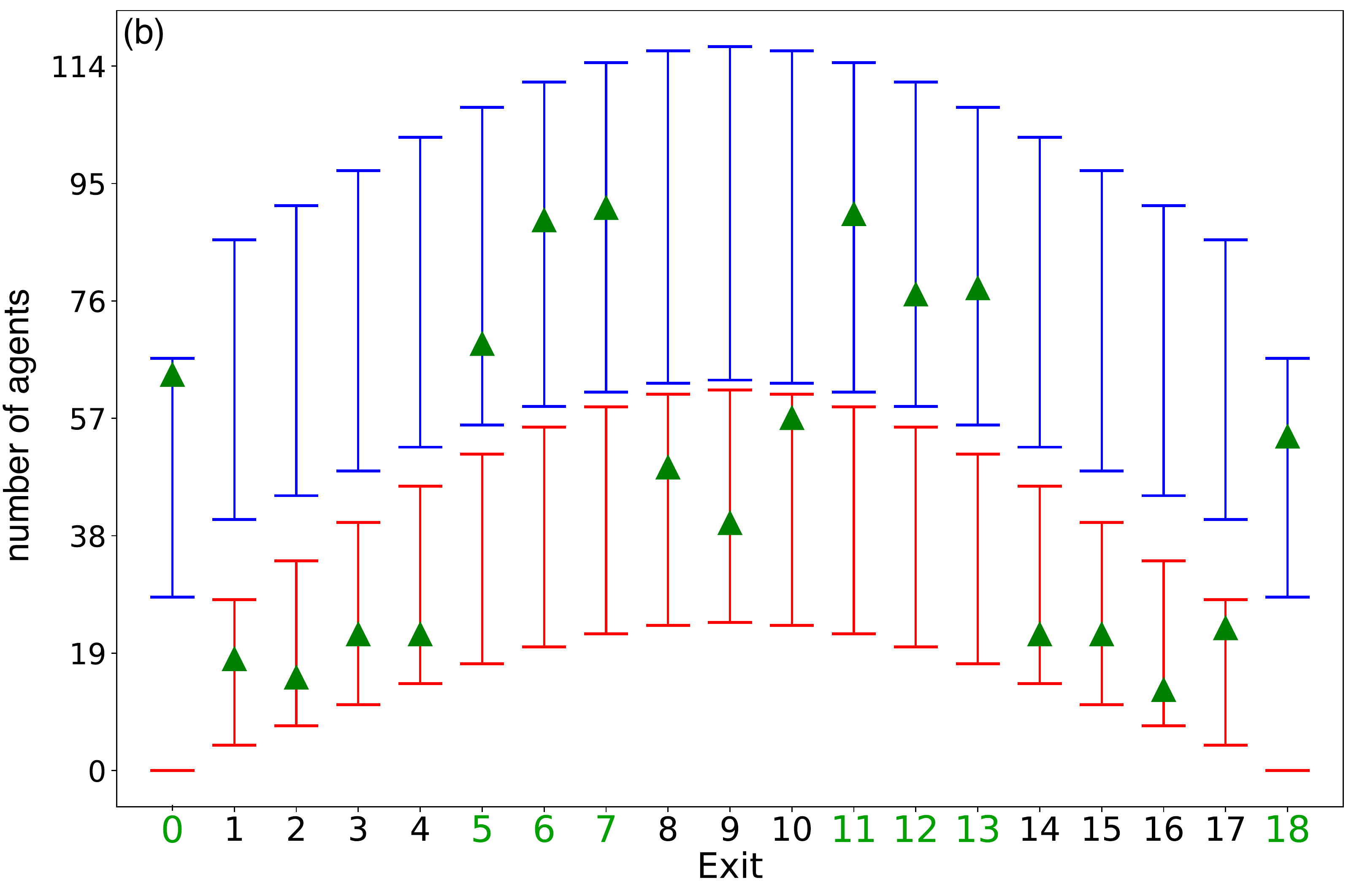}
  \includegraphics[width=0.49\textwidth]{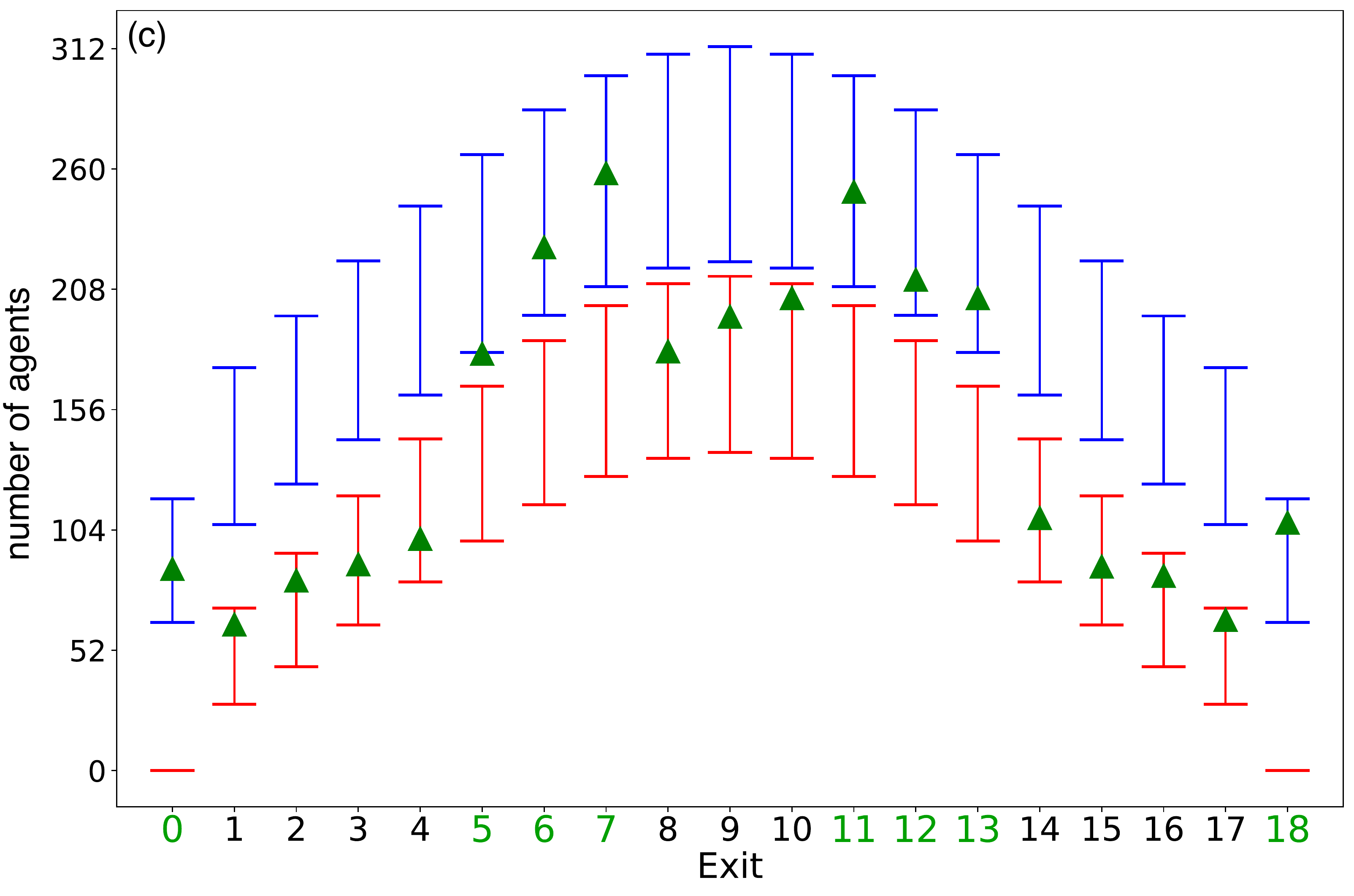}
\includegraphics[width=0.49\textwidth]{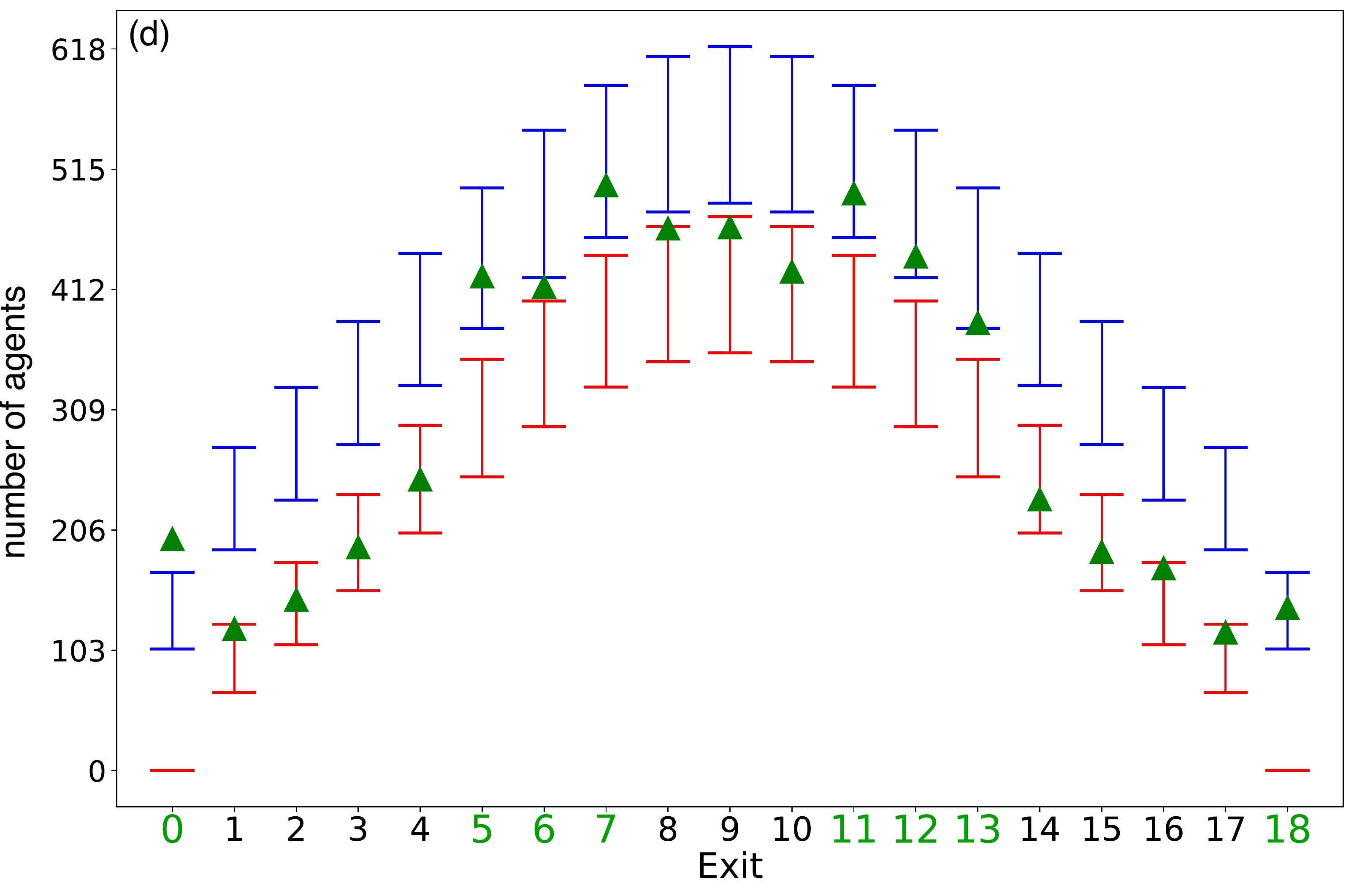}
\caption{Number of agents leaving a nonideal grid for the set  $\{5,6,7\}$ and increasing  pass junction error probability (a) $p_\text{PJ} =0.02$, (b) $ p_\text{PJ} =0.05$,  (c) $p_\text{PJ} =0.08$ and (d)  $p_\text{PJ} =0.1$. A large  error probability  leads to a stronger curvature of the confidence intervals. The minimal number of agents significantly increases from (a) $N_\text{min}^\text{non} = 287$,  (b) $N_\text{min}^\text{non} = 915$, (c) $N_\text{min}^\text{non}= 2785$ to (d) $N_\text{min}^\text{non} = 5765$. Effectively distinguishing faulty and correct agents becomes increasingly difficult with larger error probability $p_\text{PJ}$, {see, for instance, exits 6 and 8 in (d)}.  Triangles furthermore move to the edges of the respective confidence intervals, indicating the limitations of the approximation used.}
\label{3ersetvgl}
\end{figure} 

\noindent \textbf{\large 4. Comparison with  numerical simulations}\\

  \noindent
In this Section, we compare the above approximate theoretical results with exact numerical simulations of the problem for different set sizes and error probabilities in order to test their validity. Simulations are performed by {reproducing the stochastic motion of biological  agents through the grid using random numbers: at  each step, agents randomly choose to either propagate along the same direction or change direction, with a given probability.}

We begin by comparing in figure~\ref{errornumericsvgl}  the theoretical probability distribution of error-induced outcomes $p_i^\text{non}$, \eqref{nonprob},  (orange squares) for {purely} faulty pass junctions in a simplistic grid of length 18,  consisting only of pass junctions, and the corresponding numerical simulations (green triangles),  for (a) $p_\text{PJ}=0.01$ and (b) $p_\text{PJ}=0.15$. 
We observe a flat distribution for small error probabilities, indicating that outcomes are equiprobable. For larger error probabilities, the distribution is peaked at the position of the central exit point  and outcomes at the edges of the grid are less likely. The distribution is symmetric around  the center exit point 9. We have excellent agreement between theory and simulations in both cases.
  
Figure~\ref{errornumericsvglsplit} displays the analytical probability distribution of error-induced outcomes $p_i^\text{non}$, \eqref{nonprob}, for a grid consisting of pass and split junctions, neglecting the effect of split junctions  (red circles), using the approximate averaging of the split junction effect as described by \eqref{nocheinp} (orange squares), and the true error probability from the simulations by discarding the correct paths (blue diamonds), for (a),(c) $p_\text{PJ}=0.01$ and (b),(d) $p_\text{PJ}=0.15$, for the sets $\{5,6,7\}$ (top) and $\{2,3,5,7\}$ (bottom). The incorporation of split junctions  leads to  distributions peaked  at the center for all error probabilities and set sizes. The approximate treatment of the interplay between the erroneous pass and split junctions via the effective probability \eqref{nocheinp} exhibits very good agreement with the exact simulations for larger error probabilities and larger set sizes. This is due to the effect that higher error probabilities for the pass junctions as well as more pass junctions increases the number of wrong turns an agent makes while traversing through the pass junctions. This results in the fact that the split junctions can be better taken into account in the averaged treatment of \eqref{nocheinp}.
However, simulations exhibit additional inhomogeneities not fully captured by the approximations. {While for small error probabilities, the theoretical description deviates more strongly from the correct distribution, the number of wrong agents is also smaller, rather a perturbation of the correct agents. For larger error probabilities the theoretical description is closer to the true distribution, but the errors become also more dominant and thus a larger accuracy is also needed.}

Figure~\ref{1percentvgl} represents the final  positions of  agents in a nonideal grid  with an error probability $p_\text{PJ} = 0.01$, for various sets: $(a)$ $\{2,3,7\}$,  (b) $ \{5,6,7\}$, (c) $ \{2,3,6,7\}$  and  (d) $ \{2,3,7,6,4\}$. Compared with the ideal grid shown in Fig.~2, we here have (red) confidence intervals for wrong paths. The success probability $p_\text{PJ}^c$ of agents traversing correctly and the minimal number of required agents $N_\text{min}^\text{non}$, \eqref{4}, both depend on the size of the sets as well as on the number of pass and split junctions in the grid (which grows with the magnitude of the elements $N_i$ of a set). The success probability decreases with size and number of junctions: from (a) $p_\text{PJ}^\text{c} = 0.91$, (b) $p_\text{PJ}^\text{c} = 0.86$, (c) $p_\text{PJ}^\text{c} = 0.87$ to (d) $p_\text{PJ}^\text{c}= 0.84$, whereas the minimal number of agents increases from (a) $N_\text{min}^\text{non} = 169$,  (b) $N_\text{min}^\text{non} = 182$, (c) $N_\text{min}^\text{non}= 474$ to (d) $N_\text{min}^\text{non} = 1350$. A comparison with the ideal case yields the following ratios, (a)$ N_\text{min}^\text{non}/N_\text{min}= 2.14$, (b) $N_\text{min}^\text{non}/N_\text{min} = 2.3$, (c) $N_\text{min}^\text{non}/N_\text{min} = 2.86$, and (d) $N_\text{min}^\text{non}/N_\text{min} = 3.98$. In addition, a larger number of pass junctions increases errors and thus the size of the confidence intervals{, as well as  the curvature of the confidence intervals for  a fixed error probability $p_\text{PJ}$}, while a large number of split junctions leads to a stronger curvature of the overall shape of the confidence intervals caused by the corresponding  larger effective probability \eqref{nocheinp}. In all four cases, agents originating from correct and incorrect paths can be clearly distinguished. {The agents exceeding the upper bound of the confidence intervals in (c) and (d) are due to exits being reached by more than one  combination and are still valid.}

We finally analyze in  figure~\ref{3ersetvgl} the effect of an increasing  pass junction error probability (a) $p_\text{PJ} =0.02$, (b) $ p_\text{PJ} =0.05$,  (c) $p_\text{PJ} =0.08$ and (d)  $p_\text{PJ} =0.1$ for the set $\{5,6,7\}$.
 The probability of an agent correctly traversing the nonideal grid   significantly decreases as the error probability increases from  (a) $p_\text{PJ}^\text{c} = 0.74$ (b) $p_\text{PJ}^\text{c}=0.46$, $p_\text{PJ}^\text{c}=0.29$ to $p_\text{PJ}^\text{c}=0.2$. The relative number of faulty agents therefore grows markedly compared to the number of correct agents. We also see that a higher pass junction error probability  causes a stronger curvature of the confidence intervals. The minimal number of agents  additionally  gets quickly bigger with increasing error probability compared to the ideal grid: (a) $N_\text{min}^\text{non}/N_\text{min} = 3.6~ (n_i = 10)$,  (b) $N_\text{min}^\text{non}/N_\text{min} = 11.6~ (n_i=28)$,  (c) $N_\text{min}^\text{non}/N_\text{min}= 36.3 (n_i = 64)$,  (d) $N_\text{min}^\text{non}/N_\text{min}  = 72.97~(n_i = 104)$.  We furthermore note that effectively distinguishing wrong and correct exits becomes increasingly difficult with larger $p_\text{PJ}$ (see, for example, exits 6 and 8 in (d)), as triangles move to the edges of the respective confidence intervals{. Therefore, one might use $p_\text{PJ}^\text{c} > 0.5$ as a limiting value above which  the used approximations become insufficient. }\\

 \noindent \textbf{\large 5. Conclusions}\\
 
 \noindent
  We have performed a detailed analytical study of the influence of imperfect pass and split junctions on the ability of a parallel biological computer to solve  the subset sum problem. In the ideal case, an error-free grid may in principle be used to correctly solve the problem for any size, as long as the number of agents is larger than the minimum number given  by \eqref{numberOfAgentsPerfect}. For a nonideal grid, our findings indicate that the subset sum problem can still be properly solved by increasing the number of agents, provided the error probability of the different junctions is small, {so that the number of erroneous agents can be seen as small compared to the correct agents.}  As a general rule, errors in pass junctions should be as small as possible, as  {their distribution gets skewed} by the existence of  split junctions, resulting in nonhomogenous noise patterns.  Errors in split junctions alone are, however, less problematic because they simply cause outcomes to become less likely and, thus, only necessitate a larger number of agents. On the other hand, larger error probabilities make the estimation of the confidence intervals and the distinction of correct and incorrect paths increasingly difficult{, requiring either more elaborate error estimations or discarding wrong agents from the computations.}
  
  \subsection*{Acknowledgments}
 This project has received funding from the European Union’s Horizon 2020 research and innovation programme under grant agreement No 732482 (Bio4Comp).

\end{document}